\documentclass[10pt]{amsart}
\usepackage{latexsym}
\usepackage{amsfonts}
\usepackage{epigraph}

\usepackage[headings]{fullpage}

\usepackage{amsmath,amsthm,amssymb}
\usepackage{epsf}
\usepackage{psfrag}
\usepackage{graphicx}

\author[Ilya Kapovich]{Ilya Kapovich}

\address{\tt Department of Mathematics, University of Illinois at
 Urbana-Champaign, 1409 West Green Street, Urbana, IL 61801, USA}
 \email{\tt kapovich@math.uiuc.edu}

\title[Musings on generic-case complexity]{Musings on generic-case complexity}

\newtheorem{thm}{Theorem}[section] 
 
 \theoremstyle{definition}
\newtheorem{defn}[thm]{Definition}
\newtheorem{notation}[thm]{Notation}
\newtheorem{conv}[thm]{Convention} \newtheorem{rem}[thm]{Remark}
\newtheorem{exmp}[thm]{Example}

\def\strutdepth{\dp\strutbox}
\def \ss{\strut\vadjust{\kern-\strutdepth \sss}}
\def \sss{\vtop to \strutdepth{
\baselineskip\strutdepth\vss\llap{$\diamondsuit\;\;$}\null}}

\def\strutdepth{\dp\strutbox}
\def \sst{\strut\vadjust{\kern-\strutdepth \ssss}}
\def \ssss{\vtop to \strutdepth{
\baselineskip\strutdepth\vss\llap{$\spadesuit\;\;$}\null}}

\def\strutdepth{\dp\strutbox}
\def \ssh{\strut\vadjust{\kern-\strutdepth \sssh}}
\def \sssh{\vtop to \strutdepth{
\baselineskip\strutdepth\vss\llap{$\heartsuit\;\;$}\null}}

\newcommand{\N}{\mathbb N}

\def\epsilon{\varepsilon}
\def\phi{\varphi}

\newcommand{\Gen}{\mbox{Gen}}

\newcommand{\SGen}{\mbox{SGen}}

\newcommand\isom{\mathrel{\text{%
   \setbox0\hbox{$\rightarrow$}%
   \rlap{\hbox to \wd0{\hss\raisebox{0.9\height}{$\sim$}\hss}}\box0
}}}

\begin{document}

\begin{abstract}
We propose a more general definition of generic-case complexity, based on using a random process for generating inputs of an algorithm and using the time needed to generate an input as a way of measuring the size of that input.
\end{abstract}

\thanks{The author was supported by the NSF
  grant DMS-1405146.}


\maketitle

\epigraph{I have committed the sin of falsely proving Poincare's Conjecture. But that was in another country; and besides, until now no one has known about it.}{John R. Stallings~\cite{St66}}

\bigskip

\section{Introduction}

Paraphrasing one of my mathematical heroes, John R. Stallings, I am committing a mathematical sin. I am writing a paper about a definition, with no theorems, corollaries, lemmas, or propositions. And yet I will try to say something that I think is worth saying.

The notion of \emph{generic-case complexity} was introduced in the 2003 paper, by myself, Alexei Myasnikov, Paul Schupp and Vladimir Shpilrain~\cite{KMSS03}. The idea was to define the notion of a complexity class that captures the behavior of an algorithm on "most" inputs of a particular problem.
The point of such a notion is to reflect practical behavior of various algorithms. The key difference with an older notion of \emph{average-case complexity} is that generic-case complexity completely ignores the possibly bad behavior of the algorithm on a "negligible" set of inputs, instead of trying to average such behavior against good behavior on typical inputs.
Our 2003 paper dealt with generic-case complexity of various group-theoretic algorithmic problems, which subsequently has become an area of active study in  in group theory; see for example~\cite{BNW08,BMR07a,BMR07b,DMW14,GMO10,K11,KSS06,MSU08,MU11,S10}.

Since then, the notion of generic-case complexity has left the confines of group theory and has been explored in much wider mathematical and computational complexity contexts; see for example~\cite{GMMU07,K11,M09,MR08,R07,R10,R11,R13}.  Recent work of Jockusch-Schupp~\cite{JS12},  Downey-Jockusch-Schupp~\cite{DJS13}, Downey-Jockusch-McNicholl-Schupp~\cite{DJMS13}, Igusa~\cite{I13}, and others, started a systematic abstract development of the theory of generic and coarse computability in the framework of recursion theory.  The ideas of generic-case complexity also found applications outside mathematics, particularly in dealing with various big data types; see for example \cite{BCMM14,H12,HSD14,MBCM14}.

An essential feature of our original definition from \cite{KMSS03} included using the concept of "asymptotic density", defined in terms of balls with respect to some sort of a "size" function on the infinite set of inputs $\Omega$, in order to define the notions of "generic" and "negligible" subsets of $\Omega$. Most subsequent more general definitions of generic-case complexity still try to retain the same basic vocabulary of using asymptotic density and balls of increasing radius in the main definitions. 

The purpose of this note is to suggest a more organic definition of generic-case complexity, which drops the language of asymptotic density and does not use balls or a size function. Instead, this alternative approach defines genericity in terms of a random process generating inputs for an algorithm,  and uses  the time needed by a random process to produce an input for measuring the size of that input. 

The main new definition, of generic-case time complexity of an algorithm with inputs from a set $\Omega$ generated by a discrete time random process \[\mathcal W=W_1,W_2,\dots, W_n,\dots\] is given in Definition~\ref{defn:gcc3}. The definition of generic-case time complexity classes with respect to $\mathcal W$ is then given in  Definition~\ref{defn:gc-class}. 
The set-up is sufficiently general to apply to a wide variety of data types as inputs of algorithms. Thus $\Omega$ may consists of words in some alphabet, graphs, complexes, labeled diagrams,  mechanical configurations, configurations of pixels, etc. 

Upon close inspection one observes that almost all the existing results in the literature regarding generic-case complexity are covered by Definition~\ref{defn:gcc3}  and  Definition~\ref{defn:gc-class}.

The paper is organized as follows. In Section~\ref{sect:orig} we recall the original definition of generic-case complexity from \cite{KMSS03} and a version of this definition that is currently standard (at least according to such an august source of wisdom and knowledge as Wikipedia).  In Section~\ref{sect:new} we discuss some drawbacks of these definitions and propose new ones, given in Definition~\ref{defn:gcc3} and Definition~\ref{defn:gc-class}. We the discuss some features of these new definitions and the various degrees of flexibility for modifying and generalizing these definitions further; see, in particular, Remark~\ref{rem:flex}. In Section~\ref{sect:special} we discuss some important special cases, particularly the classic computability theory context of $\Omega=\{0,1\}^\ast$. 
In Section~\ref{sect:drawbacks} we discuss some drawbacks and limitations of Definition~\ref{defn:gcc3} and Definition~\ref{defn:gc-class} and suggest several possible fixes for addressing these drawbacks.

I am grateful to Paul Schupp for informative and illuminating discussions about complexity theory.

\section{The original definition}\label{sect:orig}

\begin{notation}[Running time of an algorithm]\label{not:run}
Let $\mathfrak A$ be a partial deterministic algorithm (in some model
of computation) with inputs from a set $\Omega$ and outputs in some
set $U$. For an input $w\in \Omega$ denote by $t_\mathfrak A(w)$ the
running time of $\mathfrak A$ on the input $w$ to compute an output in
$U$. Namely, $t_\mathfrak
A(w)=n\in \{1,2, 3,\dots, \}$ if $\mathfrak A$, starting with input
$w$, terminates in $n$ steps and outputs a value in $U$, and $t_\mathfrak
A(w)=\infty$ otherwise. Thus $t_\mathfrak A(w)=\infty$ precisely when,
starting on input $w$, either $\mathfrak A$ runs forever or $\mathfrak
A$ terminates in finitely many steps but outputs no value in $U$. 
\end{notation}

We recall a simplified version of the original definition used in~\cite{KMSS03}.

\begin{conv}[Basic assumption.]\label{conv:basic} For the remainder of
  this section, unless specified otherwise,  we make the following assumption. Let $\Omega$ be a countably infinite set  (where $\Omega$ is understood to be the set of all possible inputs for a particular algorithmic problem). Let $|.|: \Omega\to \mathbb Z_{\ge 0}$ be a \emph{size function}. For an integer $n\ge 0$ denote by $B_\Omega(n)$ the set of all $w\in \Omega$ with $|w|=n$. Assume that for all $n\ge 0$ the set $B_\Omega(n)$ is finite.
\end{conv}

\begin{defn}\label{defn:gen1}
For a subset $S\subseteq \Omega$ define the \emph{lower asymptotic density} $\underline{\rho}_\Omega(S)$ of $S$ in $\Omega$ as
\[
\underline{\rho}_\Omega(S)=\limsup_{n\to\infty} \frac{\#(B_\Omega(n)\cap S)}{\#(B_\Omega(n))}.\tag{$\dag$}
\]
If in the above formula the actual limit of the sequence $\frac{\#(B_\Omega(n)\cap S)}{\#(B_\Omega(n))}$ exists, we call this limit the \emph{asymptotic density} of of $S$ in $\Omega$ and denote it $\rho_\Omega(S)$.  A subset $S\subseteq \Omega$ is called \emph{generic} in $\Omega$ if $\rho_\Omega(S)=1$ (which is equivalent to the condition that  $\underline{\rho}_\Omega(S)=1$). 
A  subset $S\subseteq \Omega$ is called \emph{exponentially generic} in $\Omega$ if  $\lim_{n\to\infty} \frac{\#(B_\Omega(n)\cap S)}{\#(B_\Omega(n))}=1$ and the convergence in this limit is exponentially fast.
\end{defn}

In~\cite{KMSS03} the main cases we considered were where $\Omega$ is either the set of all freely reduced words or the set of all cyclically reduced words over some group alphabet.

For the purposes of current discussion the reader should concentrate on the case $\Omega=A^\ast$, where $A=\{a_1,\dots, a_m\}$ is a finite alphabet with $m\ge 2$ letters, and where for $w\in A^\ast$ the size $|w|$ is the length of the word $w$. 
In this case $\#(B_\Omega(n))=1+m+m^2+\dots +m^n=\frac{m^{n+1}-1}{m-1}$, and grows roughly as $m^n$ as $n\to\infty$.

Generic-case complexity is then defined as follows.
\begin{defn}\label{defn:gcc1}
Let $\mathfrak A$ be a partial deterministic algorithm with inputs
from the set $\Omega$ and outputs in some countable set $U$.

For a monotone-nondecreasing function $f(n)\ge 0$ we say that $\mathfrak A$ has \emph{generic-case time complexity} $\le f$ (correspondingly,  \emph{strong generic-case time complexity} $\le f$) if there exists a generic subset $S\subseteq \Omega$ (correspondingly, an exponentially generic subset $S\subseteq \Omega$), such that for every $w\in \Omega$ we have $t_\mathfrak A(w) \le f(|w|)$.
\end{defn}

\begin{rem}\label{rem:model}
Strictly speaking, Definition~\ref{defn:gcc1} is only a meta-definition. To make this definition formally precise we need to fix a mathematical model of computation where there is a well-defined notion of a partial deterministic algorithm with inputs from the set $\Omega$. The first, and most frequently used, way of addressing this issue is to fix an encoding of elements of $\Omega$ by finite binary sequences (that is, natural numbers) or strings in some fixed finite alphabet,  and then use the classic definition of a partially computable function (based on Turing machines).
Alternatively, one can choose another model of computation with a notion of an algorithm with inputs from the set $\Omega$ that is based on a computational device other than a Turing machine. 

We will assume that some such choice is made before applying Definition~\ref{defn:gcc1}. In Convention~\ref{conv:model} in Section~\ref{sect:new} below, before proposing a new definition of generic-case complexity, we explicitly assume choosing a particular model of computation and drop the requirement for the set $\Omega$ to be countable.
\end{rem}

Several limitations of Definition~\ref{defn:gcc1} quickly became apparent. Definition~\ref{defn:gcc1} is supposed to capture the practical behavior of $\mathfrak A$ on "typical" inputs $\omega\in \Omega$ and implicitly assumes that, for a large $n\ge 1$, a typical input $w\in B_\Omega(n)$ corresponds to the uniform probability distribution on the finite set $B_\Omega(n)$ . However, as a practical matter, even in very nice situations, there is often no easy \textbf{practical} way of choosing uniformly at random an element $w\in B_\Omega(n)$. Even in cases where such a practical procedure exists, the inputs for $\mathfrak A$ may be supplied by a random process which generates a probability distribution on  $B_\Omega(n)$ which is far from uniform.

In fact, even in the model case of $A=\{a_1,a_2\}$ and $\Omega=A^\ast=\{a_1,a_2\}^\ast$, the "natural" practical random process is given by a sequence of i.i.d random variables $X_1,X_2, \dots, X_n,\dots$ with $X_i\in A$ having the uniform probability distribution on $A$. After $n$ steps this process generates a word $W_n=X_1X_2\dots X_n\in A^n$ of length $n$ and gives a uniform probability distribution $\nu_n$ on the $n$-sphere $S(n)=A^n$ of cardinality $2^n$ in $\Omega$.  To get a uniform probability distribution on the ball $B_\Omega(n)$ of cardinality $2^{n+1}-1$, we then have to take the weighted sum $\sum_{i=0}^n \frac{2^i}{2^{n+1}-1} \nu_i$.  That is, in order to pick uniformly at random an element from the ball $B_\Omega(n)$, we first need to pick a random integer $j\in \{0,1,\dots, n\}$, where the probability of picking $i$ is $\frac{2^i}{2^{n+1}-1}$, and then pick uniformly at random an element $w$ from the $j$-sphere $A^j$. While it is possible to program an actual computer simulation of such a process, that's not what one usually does in practice when choosing elements of $A^\ast$.  Instead, one usually just runs the above mentioned sequence of i.i.d.s $X_1,X_2, \dots, X_n$ for $n$ steps and by concatenating generates a uniformly random word $W_n$ of length $n$. That is, one works with a random process that after $n$-steps generates a probability distribution on the ball  $B_\Omega(n)$ whose support is actually just the $n$-sphere $S(n)=A^n$. Thus in the nice setup of $\Omega=A^\ast=\{a_1,a_2\}^\ast$ working with uniform probability distributions on $n$-spheres is more natural than with uniform probability distributions on balls, although in this case both of these approaches produce the same notion of a generic subset of $A^\ast$ (see \cite{KMSS03} for details). 

More importantly, in most other situations, various natural random processes generating inputs in some set $\Omega$ as in Convention~\ref{conv:basic}, result in distributions on $B_\Omega(n)$ that are far from uniform. This happens, for example, for random walks on groups and graphs, various random processes generating planar diagrams, higher dimensional simplicial or cell complexes, etc.
For example, if $\Gamma$ is a finite connected graph, then a simple random walk of length $n$ on $\Gamma$ starting at a particular vertex $v_0$ can be used to generate paths of length $n$ starting at $x_0$. However, unless the graph happens to be highly symmetric, the distribution on the set of all paths of length $n$ in $\Gamma$ starting at $x_0$ determined by this random walk will be far from uniform. Moreover, there will be no  alternative practical way of picking such a path uniformly at random.

The standard way of addressing this issue is captured by the following generalization of Definition~\ref{defn:gen1}:

\begin{defn}\label{defn:gen2}
Let $\Omega$ and $|.|$ be as in Convention~\ref{conv:basic}. Let $\mu_n$, where $n=1,2,\dots,$  be a sequence of probability distributions on $B_\Omega(n)$.
A subset $S\subseteq \Omega$ is \emph{generic} for $(\mu_n)_n$ if $\lim_{n\to\infty}\mu_n(S\cap B_\Omega(n))=1$. If, moreover, the convergence in this limit is exponentially fast, then $S$ is said to be \emph{exponentially generic} for $(\mu_n)_n$. 
\end{defn}

One can then modify Definition~\ref{defn:gcc1} by using Definition~\ref{defn:gen2} instead of Definition~\ref{defn:gen1} to get a more general notion of generic-case time complexity, with respect to $(\mu_n)_n$, for an algorithm $\mathfrak A$:

\begin{defn}\label{defn:gcc1'}
Let $\Omega$ and $|.|$ be as in Convention~\ref{conv:basic}. Let $\mu_n$, where $n=1,2,\dots,$, be a sequence of probability distributions on $B_\Omega(n)$.

Let $\mathfrak A$ be a partial deterministic algorithm with inputs from the set $\Omega$ and outputs in some countable set $U$. 

For a monotone-nondecreasing function $f(n)\ge 0$ we say that $\mathfrak A$ has \emph{generic-case time complexity} $\le f$ (correspondingly,  \emph{strong generic-case time complexity} $\le f$) with respect to $(\mu_n)_n$  if there exists a subset $S\subseteq \Omega$,  which is generic for $(\mu_n)_n$ (correspondingly, exponentially generic for $(\mu_n)_n$) and such that for every $w\in \Omega$ we have $t_\mathfrak A(w) \le f(|w|)$.

\end{defn}

\section{A more organic approach}\label{sect:new}

Definition~\ref{defn:gen2}  is technically sufficient to account for a wide variety of situations where one wants to talk about generic-case complexity. Nevertheless, Definition~\ref{defn:gen2} still has several philosophical and practical drawbacks.

First, talking about generic subsets is not the most natural thing to do in probability theory. Rather, from the probabilistic point of view, it is more natural to talk about \textbf{events}. If $X$ is a measure space with a probability measure $\mu$, saying that some event happens $\mu$-almost surely does mean that this event corresponds to a subset of $S\subseteq X$ with $\mu(S)=1$. However, most mathematical arguments in probability theory are phrased in terms of estimating or computing probabilities that certain events occur (rather than in terms of talking about such events as specific subsets of the sample space). Thinking in terms of events rather than explicitly defined subsets is crucial for making probability theory work.  

Moreover, even more crucially, if we think of $(\mu_n)_{n\ge 1}$ as measures on $\Omega$ corresponding to a sequence of "random" choices of elements of $\Omega$, the sample space $\Omega_\ast$ of such a sequence is (a subset of) $\Omega^\N$ rather than $\Omega$.  Events corresponding to such sequences of random choices of elements of $\Omega$ are subsets of $\Omega_\ast$. Thus genericity should really be understood in terms of subsets of $\Omega_\ast$ rather than of $\Omega$.  Consider, for example, the situation where $\Omega=\{a,b\}$ and where $\mathfrak A$ is a partial algorithm with inputs from $\Omega$ such that $t_\mathfrak A(a)=1$ and $\mathfrak A(b)=\infty$. Let $\mathcal W=W_1,W_2,\dots$ be a sequence of i.i.d. $\Omega$-valued random variables where each $W_i$ having the uniform probability distribution $\mu$ on $\Omega$ (that is, $\mu(a)=\mu(b)=1/2$.)  Then, by the Law of Large Numbers, with probability tending to $1$ as $n\to\infty$ (that is, "generically"), for a sequence $W_1,\dots, W_n$ generated by $\mathcal W$ the algorithm $\mathfrak A$ terminates in a single step on at least $n/3$ of the inputs $W_1,\dots, W_n$. Yet, it is impossible to express this statement in terms of genericity of subsets of $\Omega$ itself.

Second, the insistence on the fact that $\mu_n$ be supported on the $n$-ball $B_\Omega(n)$ with respect to some specific "size function" $|.|$ is not always natural and ultimately unnecessary.
For various kinds of "growth" random processes, generating planar graphs, van Kampen diagrams, complexes, etc, there is not necessarily a clear choice of a specific size function $|.|$on $\Omega$ (and often several possible choices make sense).

To rectify these issues we propose Definition~\ref{defn:gcc3} below.
Before stating this definition, we adopt the following convention:

\begin{conv}\label{conv:model}
Fix a model of computation $\mathcal M$, which involves specifying the
set $\Omega$ of all possible inputs for an algorithm and a description
of allowable computational devices, such as Turing machines,
Blum-Shub-Smale machine (for computations with real numbers), etc. We
do not require $\Omega$ to be countable but, to simplify exposition,
we restrict ourselves to deterministic models of computation. We also
assume that the computational devices in $\mathcal M$ operate in
discrete time, which a single computational step taking exactly one
unit of time. As in Notation~\ref{not:run}, if  $\mathfrak A$ is a partial deterministic algorithm with inputs from the set $\Omega$,
then for $w\in \Omega$ the running time of $\mathfrak A$ on the input
$w$ to produce an output in $U$ is denoted by $t_\mathfrak A(w)$. Thus
$t_\mathfrak A(w)$
is either a positive integer or $\infty$, where the latter happens
exactly when the algorithm $\mathfrak A$ starting on the input $w$
either runs forever or terminates in finite time without producing an
output in $U$.
\end{conv} 

Our main definition is:

\begin{defn}[Generic-case complexity of an algorithm with respect to a random process]\label{defn:gcc3}

Let $\mathfrak A$ be a partial deterministic algorithm with inputs from the set $\Omega$ and values in some set $U$.

Let $\mathcal W=W_1,W_2,\dots, W_n,\dots$ be a discrete time random process that, after $n$ steps, generates an input $W_n\in \Omega$.

For a monotone-nondecreasing function $f(n)\ge 0$ we say that $\mathfrak A$ has \emph{generic-case time complexity $\le f$ with respect to $\mathcal W$}  if 

\[
\lim_{n\to\infty} Pr(t_\mathfrak A(W_n)\le f(n))=1. \tag{$\ddag$}
\]
If, moreover, the convergence in this limit is exponentially fast, we say that $\mathfrak A$ has \emph{strong generic-case time complexity $\le f$ with respect to $\mathcal W$}.
\end{defn}

In the above definition we take the view that the time $n$, needed to generate an input $W_n$ after $n$ steps of $\mathcal W$, serves as a reasonable way of measuring the size of $W_n$. As a practical matter, for many computationally natural random processes $\mathcal W$ there is (often more than one) choice of a size function $|.|$ such that we always have  $|W_n|\le O(n)$ or perhaps $|W_n|\le O(n\log n)$, or something similar. However, explicitly specifying such a size function and making it a part of the definition of generic-case complexity is not really necessary.

Definition~\ref{defn:gcc3}  makes it unnecessary to first define the notion of a generic subset of $\Omega$ and phrases condition $(\ddag)$ in terms of probabilities of events rather than in terms of existence of generic sets.  Formally $(W_n)_{n\ge 1}$ is a sequence of $\Omega$-valued random variables, so that for each $n\ge 1$ $W_n$ gives a probability distribution $\mu_n$ on $\Omega$.

In practice one would want to concentrate on the situation where each $\mu_n$ is finitely supported, and where the random process $\mathcal W=W_1,W_2,\dots, W_n,\dots$ can be relatively easily programmed on a computer. Moreover,  Definition~\ref{defn:gcc3}  best reflects the idea of a "practical" behavior of an algorithm in the case where $\mathcal W$ is a Markov process, and where we use the input $W_n$ to "construct" the next input $W_{n+1}$.  However, formally, we do not need to impose these requirements as a part of the above definition.

Definition~\ref{defn:gcc3} also makes it more conceptually clear that the notion of generic-case complexity does depend (and crucially so) on the choice of a random process $\mathcal W$ generating the inputs of a problem. This key point often easily gets lost in the contexts of discussing versions of generic-case complexity based on asymptotic density considerations.

 One can then define the notion of a generic complexity class.
 
 \begin{defn}[Generic-case complexity class]\label{defn:gc-class}
 Let $\mathcal C$ be a deterministic time complexity class (such as linear time, quadratic time, polynomial time, exponential time, etc) for our computational model $\mathcal M$. Let  $h: \Omega\to U$ be a function, where  $U$ is another countable set. Let $\mathcal W=W_1,W_2,\dots, W_n,\dots$ be a discrete time random process that, after $n$ steps, generates an input $W_n\in \Omega$.
 
 (1)  Let $\mathfrak A$ be a deterministic partial algorithm with inputs from the set $\Omega$ and values in $U$ such that $\mathfrak A$ is \emph{correct} for $h$, that is whenever $\mathfrak A$ actually outputs some value on $w\in \Omega$, that value is equal to $h(w)$.
 
 We say that $\mathfrak A$ \emph{computes $h$ with generic-case time complexity $\mathcal C$ with respect to $\mathcal W$} if there exists a monotone non-decreasing function $f(n)\ge 0$ satisfying the time constraints of $\mathcal C$ such that $\mathfrak A$ has generic-case time complexity $\le f$ with respect to $\mathcal W$.
 
(2) We say that $h$ is \emph{computable with generic-case time complexity $\mathcal C$ with respect to $\mathcal W$} if there exists a correct partial algorithm $\mathfrak A$ for $h$ such that $\mathfrak A$ computes $h$ with generic-case time complexity $\mathcal C$ with respect to $\mathcal W$

 (3) For given $\Omega, \mathcal W$ and $\mathcal C$ we denote by $\Gen_\mathcal W(\mathcal C)$ the set of all $h: \Omega\to U$ that are computable with generic-case time complexity $\mathcal C$ with respect to $\mathcal W$.  The notion of a function $h$ \emph{computable with strong generic-case time complexity $\mathcal C$ with respect to $\mathcal W$} and the corresponding complexity class $\SGen_\mathcal W(\mathcal C)$ are defined similarly.

 \end{defn}
 
\begin{rem}\label{rem:flex}  We again stress that Definition~\ref{defn:gcc3}  and Definition~\ref{defn:gc-class} do require first choosing a model of computation $\mathcal M$, as specified in Convention~\ref{conv:model}.

Definition~\ref{defn:gcc3}  and Definition~\ref{defn:gc-class} are flexible enough to allow for a straightforward modification of generic-case complexity $\mathcal C$ with respect to $\mathcal W$, where $\mathcal C$ is a deterministic complexity class with a resource bound on space (such as log-space, linear space, etc) or on combination of time and space. One can also easily modify these definitions to allow for dealing with nondeterministic algorithms and complexity classes. If $\mathfrak A$ is a nondeterministic algorithm with inputs from $\Omega$, one just needs to define $t_\mathfrak A(w)$ as the shortest length of a computational path for $\mathfrak A$ that starts with an input $w$. 

Other variations of these definitions are possible. In particular, one can relax the assumption that the limit in $(\ddag)$ be equal to $1$ and instead require this limit (or the corresponding liminf) to be positive.  One can also relax the requirement  in part (2) of Definition~\ref{defn:gc-class} that the algorithm $\mathfrak A$ be correct for $h$, and instead require that $\mathfrak A$ produce correct values of $h$ with asymptotically positive probability or probability tending to $1$ as $n\to\infty$.  These two types of generalizations were introduced by Jockusch-Schupp~\cite{JS12} in the model context of $\Omega=\{0,1\}^\ast$ and $W_n$ picking uniformly at random a binary string of length $n$, leading to the notions of "generic computability at density $d$" (where $0<d\le 1$) and of "coarse computability".

Also, it is fairly straightforward to adapt  Definition~\ref{defn:gcc3}  and Definition~\ref{defn:gc-class} to deal with continuous-time random-processes $\mathcal W=(W_t)_{t\ge 0}$ and to drop the requirement for $\Omega$ to be countable. One just needs to work in a computational model with a well-defined notion of an algorithm (deterministic or nondeterministic) with inputs from $\Omega$.
\end{rem}
 
For Definition~\ref{defn:gcc3}  and Definition~\ref{defn:gc-class}, the case where $U=\{0,1\}$ and the function $h:\Omega\to\{0,1\}$ is the characteristic function of some subset $D\subseteq \Omega$, corresponds to the \emph{decision problem} of determining whether an element $w$ of $\Omega$ belongs to $D$. The definition of generic-case complexity considered in \cite{KMSS03} was limited to considering decision problems.

 We stress that the dependence of Definition~\ref{defn:gcc3}  and Definition~\ref{defn:gc-class} on the random process $\mathcal W$ is an essential feature of these notions which reflects the fact that the practical behavior of various algorithms \textbf{does depend} on the choice of random processes used to generate inputs for an algorithm. 
 
 We demonstrate this point on the following simple example.
 
 \begin{exmp}
 Let $G$ be a finitely generated group that splits as an HNN-extension of another finitely generated group $H$ with stable letter $t$ and associated isomorphic subgroups $L_1,L_2\le H$ ,  so that $G=\langle H, t| t^{-1}L_1t=L_2\rangle$. Let $\{b_1,\dots, b_m\}$ be a finite generating set for $H$ (where $m\ge 2$), let $X=\{t,b_1,\dots, b_m\}^{\pm 1}$,  and let $B=\{b_1,\dots, b_m\}^{\pm 1}$.  Let $\Omega=X^\ast$ be the set of all words over $X$ and let $D\subseteq X^\ast$ be the set of all words $w$ over $X$ such that $w=_G 1$. 
 
 Let $\mathcal W$ be the random process such that at time $n$ $W_n\in A^n$ is a word over $X$ of length $n$ chosen uniformly at random.  Let $\mathcal W'$ be the random process such that at time $n$ $W_n'\in B^n$ is a word over $B$ of length $n$ chosen uniformly at random.
 
 Let $\mathfrak A$ be the partial algorithm with inputs from $\Omega=X^\ast$ which proceeds as follows.  Given a word $w\in X^\ast$, the algorithm first computes the exponent sum $\sigma_t(w)$ on $t$ in $w$. If $\sigma_t(w)\ne 0$, the algorithm declares that $w\not\in D$ and terminates. If $\sigma_t(w)= 0$, the algorithm $\mathfrak A$ runs forever.  Note that for every $w\in B^\ast$ we have $t_\mathfrak A(w)=\infty$.
 
 Then, as explained in \cite{KMSS03}, $\mathfrak A$ solves the decision problem of belonging to $D$ with linear time generic-case complexity with respect to $\mathcal W$.
 On there other hand, there is no time complexity class $\mathcal C$ such that $\mathfrak A$ solves the decision problem of belonging to $D$ with  generic-case complexity in $\mathcal C$ with respect to $\mathcal W'$.
 
 \end{exmp}
 
Note that, as a practical matter,  in the above example both $\mathcal W$ and $\mathcal W'$ are equally valid ways of generating inputs in $\Omega=A^\ast$ because both these processes can be easily programmed with a computer and implemented in practice.

 Definition~\ref{defn:gcc3}  fits well with the notion of a "random" element $w\in\Omega$ having some particular property.  Thus let $D\subseteq \Omega$ be the set of all elements satisfying some property (in which case we also refer to $D\subseteq \Omega$ as a \emph{property} of elements of $\Omega$). Let $\mathcal W=W_1,W_2,\dots, W_n,\dots$ be a discrete time random process that, after $n$ steps, generates an input $W_n\in \Omega$. We say that an element of $\Omega$ \emph{has property $D$ generically with respect to $\mathcal W$} if
 \[
 \lim_{n\to\infty} Pr(W_n\in D)=1.
 \]

 \section{Important special cases}\label{sect:special}

There are several situations where there is one especially natural choice of a random process $\mathcal W$ generating elements of $\Omega$, and it makes sense to fix that choice.
In particular, if $A=\{a_1,\dots, a_m\}$ is a finite alphabet with $m\ge 2$ letters and $\Omega=A^\ast$, then it is particularly natural to use $\mathcal W=W_1,W_2,\dots, W_n,\dots$ where $W_n$ picks uniformly at random a word $w\in A^n$ of length $n$ over $A$. Thus $W_n$ induces the uniform probability distribution $\mu_n$ on $A^n$, and, as noted earlier, it is easy to practically implement the process $\mathcal W$ by using a concatenation of $n$ i.i.d. random variables with uniform probability distribution on $A$. 
Similarly, in the context of group-theoretic decision problems originally considered in \cite{KMSS03}, we are dealing with a finitely generating group $G$ with a finite generating set $A=\{a_1,\dots, a_m\}$. The inputs for such decision problems are freely reduced words in the group alphabet $X=\{a_1,\dots, a_m\}^{\pm 1}$, so that $\Omega=F(a_1,\dots,a_m)$. It is then particularly natural to use the random process  $\mathcal W=W_1,W_2,\dots, W_n,\dots$ where $W_n$ picks uniformly at random a freely reduced word of length $n$ over $X$. Again, $\mathcal W$ is easy to implement in practice by considering the simple non-backtracking random walk on $F(a_1,\dots,a_m)$. In both of the above situations the original asymptotic density approach from Definition~\ref{defn:gen1} produces the notions of genericity and generic-case complexity equivalent to those provided by Definition~\ref{defn:gcc3}  and Definition~\ref{defn:gc-class}. The reason comes from Stolz' theorem which explains why in these cases computing asymptotic density via balls of radius $n$ yields the same notion of a generic set as when using uniform probability distributions on spheres. See Lemma~5.5 and Lemma~5.6 in \cite{KMSS03} for details. 

A particularly important instance of the above situation is the case where $A=\{0,1\}$ and where $\Omega=A^\ast\setminus\{\epsilon\}$ is the set of all finite nonempty binary sequences, which is naturally identified, via binary expansion, with the set $\mathbb Z_{\ge 0}$. Then again it is especially natural to consider the random process $\mathcal W=W_1,W_2,\dots, W_n,\dots$ where $W_n$ picks uniformly at random a binary sequence $w\in \{0,1\}^n$ of length $n$. Using this $\Omega$ and this choice of $\mathcal W$ and studying in depth and detail the corresponding generic complexity classes $\Gen_\mathcal W(\mathcal C)$ provides an ideal model setting for studying the concepts provided by Definition~\ref{defn:gcc3}  and Definition~\ref{defn:gc-class} (and, as noted above, in this situation these definitions provide the same notions of generic-case complexity as those given by Definition~\ref{defn:gcc1}).  This point of view is taken in the recent work of Jockusch-Schupp~\cite{JS12},  Downey-Jockusch-Schupp~\cite{DJS13} and  Downey-Jockusch-McNicholl-Schupp~\cite{DJMS13} where a systematic development of the theory of generic and coarse computability is pursued in this setting.  "Coarse computability" refers to relaxing the requirement in Definition~\ref{defn:gc-class} that the algorithm $\mathfrak A$ be correct for the function $h$, and allowing $\mathfrak A$ to produce incorrect answers with small probability.

\section{Limitations of the new definition and possible fixes}\label{sect:drawbacks}

Definition~\ref{defn:gcc3} is based on the premise that  for each  $n=1,2,\dots$ the random process $\mathcal W$ generates an element $W_n\in \Omega$ that can be used as an input for the algorithm $\mathfrak A$, and that $n$ serves as a reasonable measure of the "size" of $W_n$.  This assumption works well in many natural situations, where $\mathcal W$ proceeds by performing "bounded local perturbations".  By that we mean that during the $n$-th step of the process the configuration $W_{n-1}\in \Omega$ is modified by some sort of a bounded local change to produce a new configuration $W_{n}\in \Omega$, which again constitutes a valid input for $\mathfrak A$. 

However, there are many situations where such an assumption about the nature of the random process $\mathcal W$ is not reasonable. For example, it may be that after some number $n$ of steps the random process does generate a valid input $W_{n}\in \Omega$. Then it performs $k$ auxiliary processing steps (each consuming a single unit of time and each constituting a "bounded local perturbation"), so that $W_{n+1}, \dots, W_{n+k}$ are "auxiliary" objects/configurations that don't belong to $\Omega$ and that are used to produce a valid input $W_{n+k+1}\in \Omega$ which can now be fed into $\mathfrak A$.  In general, $k$ itself may depend in either random or deterministic way in the previous "valid" input $W_{n}\in\Omega$, or even of the entire trajectory $W_1,..,W_{n}$ up to time $n$. Moreover, $k$ need not be uniformly bounded above by a constant independent of $n$. 
In this situation the set-up used in Definition~\ref{defn:gcc3} is not suitable. 

In this case the random process $\mathcal W=W_1,W_2, \dots $ takes values $W_n\in \Omega'$, where $\Omega'=\Omega\sqcup \Omega^{aux}$ is a bigger set, with $\Omega^{aux}$ being the set of "auxiliary" configurations.

There are several different ways in which Definition~\ref{defn:gcc3}  can be adapted to deal with this more general set-up.

{\bf Case 1.}  Suppose there is $n_0\ge 1$ such that for every $n\ge n_0$ we have $Pr(W_n\in\Omega)>0$. We can then modify Definition~\ref{defn:gcc3}  by rephrasing it it terms of conditional probabilities and replacing $(\ddag)$ by requiring that
\[
\lim_{n\to\infty} Pr(t_\mathfrak A(W_n)\le f(n)| W_n\in \Omega)=1. \tag{$\spadesuit$}
\]

{\bf Case 2.} For many natural examples of $\mathcal W$ (such as the example with generating random graphs discussed below), there is some deterministic sequence of times ${\mathbf n}=(n_i)_i$, $1\le n_1<n_2< \dots $ (independent of the trajectory of $\mathcal W$) such that for each $i\ge 1$ we always have $W_{n_i}\in \Omega$ and for each $n\not\in\{n_1,n_2,\dots, \}$ we always have $W_n\in \Omega^{aux}$. 

In such a situation we can modify Definition~\ref{defn:gcc3} and replace $(\ddag)$ by requiring that
\[
\lim_{i\to\infty} Pr(t_\mathfrak A(W_{n_i})\le f(n_i))=1. \tag{$\clubsuit$}
\]

\begin{exmp}
Consider the situation where we are trying to generate random graphs on $n$ vertices with $n\to\infty$.
The set $\Omega$ is defined as $\Omega=\sqcup_{n=1}^\infty \Omega[n]$ where $\Omega[n]$ consists of all simple graphs with vertex set $\{1,2,\dots,n\}$.

If at some stage of the process we have constructed a graph $\Gamma\in \Omega[n]$, we then construct a graph $\Delta\in \Omega[n+1]$ as follows. We first add a new vertex $n+1$ to $\Gamma$ and then perform $n$ independent flips of a fair coin $n$ to decide whether to put an edge between vertex $n+1$ and vertices $1,\dots, n$.  Thus we used $n+1$ extra  steps to produce a sequence $\Gamma_1,\dots, \Gamma_{n+1}=\Delta$, where we view $\Gamma_{n+1}=\Delta$ as an element of $\Omega[n+1]$, and view the "auxiliary" graphs  $\Gamma_1,\dots, \Gamma_n$ as elements of $\Omega^{aux}$.  (Although it is possible to think of  $\Gamma_1,\dots, \Gamma_n$ as elements of $\Omega[n+1]$, we can formally enforce the condition $\Gamma_1,\dots, \Gamma_n\not\in\Omega[n+1]$ by adding a "flag" register to each of these graphs.)

 A reasonable choice of $\mathcal W$ here would proceed as follows. We always put $W_1$ to be the graph consisting of a single vertex.  We then start applying the above procedure iteratively.
This defines a random process $\mathcal W=W_1,W_2,\dots $ such that for $i=1,2,\dots $ and $n_i=1+2+\dots +i=i(i+1)/2$ we have $W_{n_i}\in \Omega[i]$ and for each $n\not\in\{n_1,n_2,\dots, \}$ we always have $W_n\in \Omega^{aux}$.
 
\end{exmp}
 
 \begin{exmp} There are situations where neither $(\spadesuit)$ nor $(\clubsuit)$ gives a suitable generalization of Definition~\ref{defn:gcc3}.  For example, let $\Omega=\sqcup_{n=1}^\infty S_n$ be the disjoint union of symmetric groups $S_n$. Suppose we are trying to devise a random process that, as $n\to\infty$,  produces uniform probability distributions $\mu_n$ on $S_n$. 
 For a given $n$, we can generate a uniformly random permutation $\sigma\in S_n$ in a reasonable way: we create a random re-arrangement $j_1,\dots, j_n$ of the numbers $1,\dots, n$ by first picking uniformly at random $j_1\in \{1,\dots, n\}$, then picking uniformly at random $j_2\in  \{1,\dots, n\}\setminus \{j_1\}$, then picking uniformly at random $j_3\in  \{1,\dots, n\}\setminus \{j_1,j_2\}$, and so on. Then we define $\sigma\in S_n$  as $\sigma(i)=j_i$ for $i=1,\dots, n$.  However, unlike in the example with generating random graphs given above, having chosen uniformly at random $\sigma\in S_n$ does not allow us to use $\sigma$ as a starting block for building a uniformly random element of $S_{n+1}$. Formally, we can still use the approach of Case~2 given by $(\clubsuit)$ if, after having chosen a random $\sigma\in S_n$, we forget this choice completely and start building a random permutation in $S_{n+1}$ from scratch, using the above procedure. The method of Case~2 is applicable since the number of steps needed to generate a random element of $S_n$ from scratch depends only on $n$.  However, applying the approach of Case~2 in this situation does seem fairly artificial, and it may be better to use (a version of) Definition~\ref{defn:gcc1'} directly in this case. 
 \end{exmp}


\end{document}